\documentclass{iau}
\usepackage{graphicx}

\title[Planetary Evaporation and the Dynamics of Planet Wind/Stellar Wind Bow Shocks] %% give here short title %%
{Planetary Evaporation and the Dynamics of Planet Wind/Stellar Wind Bow Shocks}

\author[A. Frank, B. Liu, J. Carroll-Nellenback, A. C. Quillen, J. F. Kasting, I. Dobbs-Dixon \& E.G. Blackman]   %% give here short author list %%
{A. Frank$^1$, B. Lui,$^1$ J. Carroll-Netterback$^1$, A. Quillen$^1$, E. Blackman$^1$, J. Kasting$^2$ \& I. Dobbs-Dixon$^3$}

\affiliation{$^1$University of Rochester\\
$^2$Penn State University\\ $^2$ New York University Abu Dhabi}

\pubyear{2015}
\volume{314}  %% insert here IAU Symposium No.
\pagerange{119--126}
\jname{Young Stars \& Planets Near the Sun}
\editors{J. H. Kastner, B. Stelzer, \& S. A. Metchev, eds.}
\begin{document}

\maketitle

\begin{abstract}
We present initial results of a new campaign of simulations focusing on the interaction of planetary winds with stellar environments using Adaptive Mesh Refinement methods. We have confirmed the results of Stone \& Proga 2009 that an azimuthal flow structure is created in the planetary wind due to day/night temperatures differences.  We show that a backflow towards the planet will occur with a strength that depends on the escape parameter.  When a stellar outflow is included, we see unstable bow waves forming through the outflow's interaction with the planetary wind. 

\end{abstract}

\firstsection % if your document starts with a section,
\section{Introduction}
Planetary blow-off occurs when irradiation from the central star, especially in the extreme ultraviolet (EUV), heats of the upper layers of the atmosphere to produce an extended envelope of gas which transitions into an wind.  Such flows are believed to occur in Hot Jupiters  (i.e. HD 209458b, Ballister et al 2007).  A characteristic measure of the strength of the wind is the ratio of gravitational potential to thermal energy at the top of the atmosphere.  This is usually called the hydrodynamic escape parameter,

\begin{equation}				
\lambda=\frac{GM_p \mu}{(R_p kT_p)}
\end{equation}
\vspace{0.1cm}

where $M_p$ and $R_p$  are the mass and radius of the planet, and $T_p$ and $\mu$ are the temperature and mean mass per particle in the atmosphere. For $\lambda  >> 10$, the atmosphere is too tightly bound for a hydrodynamic wind to form.  Note that weaker outflows may be produced via non-thermal processes, e.g., Hunten (1982). For $\lambda  \sim 10$, a Parker-type, thermally driven hydrodynamic wind is expected (note that $\lambda  \sim 15$ for the sun with its $T \sim 10^6$ K corona).  Note also that the composition of the atmosphere must also be considered in detailed models of wind launching. 

When these atmospheric winds occur, they are expected to have a wide range of applications and observational consequences in exoplanet studies.  In particular many open questions exist about how these winds will change when additional physics are added.   Physical processes that could affect the wind structure and outflow rates include planetary magnetic fields (Owen and Adams 2014), time dependent EUV flux (Lecavelier des Etangs et al. 2012), atmospheric circulation (Teyssandier et al. 2015) and the interaction between stellar and planet winds (Murray-Clay et al. 2009, Stone \& Proga 2009) ).  Many of these processes have only recently begun to be incorporated into existing simulations (Schneiter et al 2007, Cohen et al 2011, Bisikalo et al 2013). The recent work of Matsakos et al 2015, for example, includes a variety of processes in a 3-D MHD study that tracked the full orbital dynamics of the wind.  

In this contribution we present initial results of a new campaign of simulations focusing on the interaction of planetary winds with stellar environments using Adaptive Mesh Refinement methods.  The goal of this initial study is to explore the multi-dimensional wind structure from photo-evaporating planets and the time-dependence of planetary wind/stellar wind interactions.  

\section{Code and Set-up}

In this study we present first results of our AMR simulations of planetary wind launching and planet-wind/stellar wind interactions using the AstroBEAR code (Cunningham et al 2009).  AstroBEAR is a fully parallelized AMR MHD multi-physics code which currently includes modules for the treatment of self-gravity, ionization dynamics, chemistry, heat conduction, viscosity, resistivity and radiation transport via flux-limited diffusion (or plane-parallel ray tracing).  For these initial simulations we explore only hydrodynamic interactions in axisymmetry, (2.5D), with a polytropic equations of state (the polytropic index $\gamma$ is an input parameter and we assume isothermal conditions $\gamma= 1.01$).  

\begin{figure}[b]
\vspace*{0.1 cm}
\begin{center}
\includegraphics[width=5.5in]{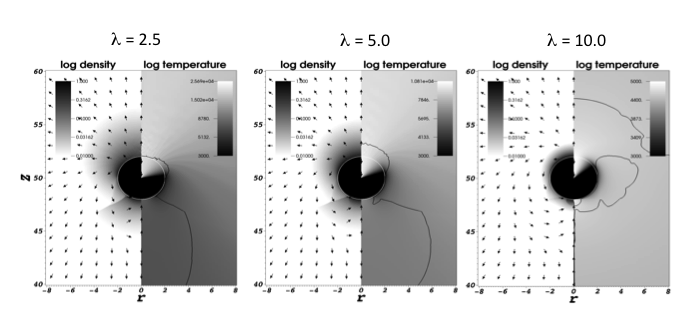} 
\vspace*{.05 cm}
\caption{Fig 1 Steady state planetary wind solutions computed at high resolution in 2.5D using AstroBEAR for different escape parameter $\lambda$ values.  Flow and density are shown on the left and thermal structure and $M=1$ contours are shown on the right.  Note the flow from the day-side (top) to the night-side passing through a shock (apparent at $\theta \sim 135^o$ from vertical) and leading to backflow on to the planetÕs night-side.}
\label{fig:grains}
\end{center}
\end{figure}

Our initial set up is similar to that of Stone \& Proga (2009) .  We initialize the outer boundary of the planet (assumed to be the wind launch point) with azimuthally variable temperature $T(\theta) = T_o sin (\theta)$ where $\theta = 90^o$ is the sub solar point and $T_o = 10^4$ K.  The high temperatures on the day side of the planet launch a strong Parker-type thermal wind.  On the night side the lower temperatures imply a lower value of $\lambda$ leading to an aspherical flow from the planet.  In some of our models we also drive a plane parallel stellar wind into the grid.  Given the small orbital radius for planets on "hot" orbits, it is possible that the planet will still be in the stellar wind acceleration region and we consider both subsonic ($M=0.8$) and supersonic ($M= 5.0$) stellar winds.

\section{Planetary Wind Structure: Nightside Backflow}
On order to asses the multi-dimensional structure of the planetary evaporative flow we have run a series of simulations at high resolution with different initial conditions for the planetary parameters.  The simulation domain has a resolution of 64 zones per planetary radii at the highest level of refinement.  We choose our planetary parameters to sample three values of the escape parameter ($\lambda = 2.5, 5, 10$).  Figure 1 shows the wind solutions (density, temperature, velocity field, Mach surface) for these runs.

In all three cases we see a strong Parker type thermal wind being launched from the day side of the planet (top hemisphere).  Consideration of the $M=1$ mach surface (black contour) demonstrates that in all 3 cases, the wind reaches supersonic velocities close the planet.  The distance to the sonic surface however depends on the escape parameter increasing with the value of $\lambda$.  The flow from the planet is not however purely radial.  As first shown by Proga \& Stone (2009), azimuthal pressure gradients in the sub-sonic and trans-sonic regions of the planetary flow drive material from the hot day side towards the colder night side of the planet.  As this flow converges towards the night side, a conical shock forms at approximately latitude $\theta=135^o$ (relative to the substellar point).  This shock weakens with increasing $\lambda$ and is not seen in our $\lambda = 10$ simulation because the flow is entirely subsonic at these radii.  We note that the converging flow on the nightside does more than just redirect outflowing material.  In our simulations we see some gas fall back towards the planetary boundary.  We find the relative strength of this backflow also decreases for the largest values of our $\lambda$ series.  

\section{Stellar Wind Interactions}
The high resolution of our simulations translates into higher Reynolds numbers and hence a greater ability to capture instabilities associated with stellar wind/planetary wind interactions.  As the stellar wind sweeps past the planet, it encounters the planetary wind and, depending on the ratio of ram pressures (or thermal pressures for subsonic flows), a bow wave may form.  Such a bow wave/bow shock may be prone to a variety of unstabilities including non-linear thin shell modes (NTSI) if the bow shock cools effectively.  Since instabilities will generate vorticity one can expect density inhomogeneities to form which will be swept downstream on a timescale of $t_s \sim R_p/V_{bs}$ where $V_{bs}$ represents the flow behind the bow shock.  

\begin{figure}[b]
\vspace*{-0.05 cm}
\begin{center}
\includegraphics[width=5.0in]{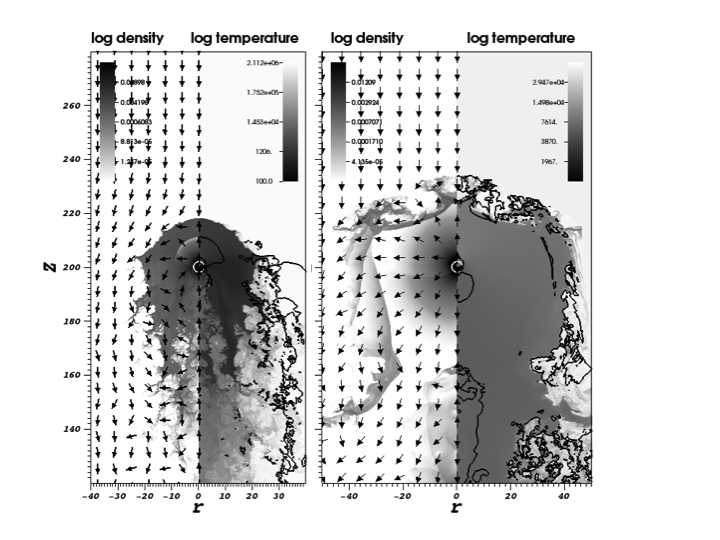} 
\vspace*{.0250 cm}
\caption{ Stellar wind/planetary wind simulations. (Left) $M_*=0.8$  and (right) $M_*=5.0$ wind flowing downward from the upper boundary to interact with a $\lambda = 5$ planetary wind.}
\label{fig:grains}
\end{center}
\end{figure}

Because the stellar wind may not have reached its final supersonic velocity we have carried out simulations representing both subsonic and supersonic stellar flows impinging on a planetary wind.  In figure 2 we show flow conditions for a $M_*=0.8$ (left) and a $M_*=5.0$ (right) wind flowing downward from the upper boundary to interact with a $\lambda = 5$ planetary wind (see figure 1).   In both the subsonic and supersonic cases we see the bow wave is not smooth but breaks up due to instabilities.  The morphology of the two cases is however quite different.  In the subsonic stellar wind case the bow consists of one shock facing back into the planetary wind.  We see that the instabilities are triggered at the contact discontinuity between the $M<1$ stellar wind and the $M>1$ planetary wind in a manner similar to the thin shell instability.  In the supersonic stellar wind case, the bow consists of two shocks facing upstream and downstream and here we see larger vortices forming in the unstable downstream flow.

\section{Conclusions}
In this contribution we have presented first results from a new campaign of AMR simulations designed to explore the interactions of an evaporative planetary wind with its stellar environment.  We have confirmed the results of Stone \& Proga (2009) showing that azimuthal structure of the wind forms due to day/night temperature differences.  Going further we have also found that such structures depend on the value of the escape parameter $\lambda$.  When a stellar outflow is included we see an unstable bow wave forming between it and the planetary wind.  The bow wave is unstable in both sub-sonic and supersonic stellar outflow cases.  The morphology of the resulting flows, however, differs in the two cases.  Thus our work implies a time-dependent nature of the interaction region (with variations expected on the crossing time $t_s \sim R_p/V_{bs}$).  Future studies will focus on carrying the simulations out in the co-rotating frame.

\end{document}